# Analysis of deviation from neoclassical ion equilibrium against electron and ion temperature profiles in T-10 tokamak

P.V. Minashin[1], A.B. Kukushkin[1,2], A.V. Melnikov[1,2], M.A. Drabinskiy[1], L.G. Eliseev[1],

P.O. Khabanov[1], N.K. Kharchev[1], S.E. Lysenko[1], M.R. Nurgaliev[1]

and T-10 team

[1]*National Research Center "Kurchatov Institute", Moscow, Russia*

[2]*National Research Nuclear University MEPhI, Moscow, Russia*

The results of the analysis of the deviation of the force equilibrium for ions from the neoclassical theory prediction, calculated using the direct measurements of the radial electric field, in the view of its possible local and nonlocal correlation with the profiles of electron, $T_e$, and ion, $T_i$, temperatures in the T-10 tokamak are presented. Local correlations are analyzed by means of the Pearson's correlation. Nonlocal correlations are treated with an inverse problem under the assumption of an integral equation relationship between the deviation and $T_e$ and $T_i$ profiles. The discharges with zero, weak and strong auxiliary heating (electron cyclotron resonance heating) are analyzed. It is found that the electrons substantially (not less than ions) contribute to the deviation of the ion equilibrium from the neoclassical theory prediction both in the local and nonlocal models.

## 1. Introduction

Measurements of the radial electric field, $E_r$, in the T-10 tokamak with HIBP diagnostics showed qualitative agreement between neoclassical theory [1]-[5] and experiment in the central part of the plasma column for a wide range of discharge parameters [6], [7]. With an increase of the electron cyclotron resonance heating (ECRH) power this agreement deteriorates [8], which correlates with the hypothesis of an increase of the anomalous electron transport with an increase of the ECRH power [9], [10]. A similar effect can be caused by an increase in anomalous electron transport with a decrease in the electron density in ohmic discharges [11].

In this work an analysis of the deviation of the predictions of the neoclassical theory from the measured values of $E_r$ in tokamak T-10 in the view of its possible local and nonlocal correlation with the profiles of electron, $T_e$, and ion, $T_i$, temperatures is carried out. Local correlations are analyzed by means of the Pearson's correlation. Nonlocal correlations are treated with an inverse problem under the assumption of an integral equation relationship between the deviation and $T_e$ and $T_i$ profiles. By solving the inverse problem, the dependence of the spatial profiles of the indicated deviation on the profiles of ion and electron temperatures is analyzed.

Deviation of the equilibrium equation for ions in T-10 tokamak (without taking into account rotation due to its relatively small contribution to the equilibrium equation) is defined as the residual (difference) between measured experimental electric field, $E_r$, and neoclassical predictions for the radial electric field in the "neoclassical" theory of MHD plasma equilibrium, $E_r^{neo}$ [1], [3] (eq. 7):

$$D(r,t) = E_r - E_r^{neo}, \tag{1}$$

$$E_r^{neo} = \frac{T_i}{e}\left(\frac{1}{n_i}\frac{\partial n_i}{\partial r} + \gamma \frac{1}{T_i}\frac{\partial T_i}{\partial r}\right), \tag{2}$$

where $r$ is the minor radius of toroidal plasma, $t$ is time, $T_i$ is the ion temperature profile and $n_i$ is the ion density profile.

The values of parameter γ in (2) should be selected in accordance with the transport mode [3].

## 2. Experimental data

Analysis is carried out for the experimental data obtained in 5 regimes of T-10 tokamak operation, which parameters are presented in Table 1 and Figure 1.

Table 1 Parameters of the T-10 tokamak regimes of operation

| T-10 regime | $\overline{n_e}$, $10^{19}$ m$^{-3}$ | $B_t$, T | $I_{pl}$, kA | $P_{EC}$, MW | $\tau_E$, ms | $r_{HIBP}$, cm | $T_e(0)$, keV | $T_i(0)$, keV | $\overline{Z_{eff}}$ | Comments |
|---|---|---|---|---|---|---|---|---|---|---|
| 1 | 0.6 | 2.1 | 150 | 0 | 14 | 10-30 | 1.3 | 0.25 | 1.7 | №61452, 650 ms |
| 2 | 2.0 | | | 0 | 49.5 | | 1.1 | 0.54 | 2.0 | №61452, 850 ms |
| 3 | 1.17 | 2.2 | 250 | 0 | 14 | 7-30 | 1.4 | 0.45 | 2.5 | №73198, 620-740 ms |
| 4 | 1.6 | | | 0.5 | 10.4 | | 1.75 | 0.6 | 2.1 | №73198, 915-950 ms |
| 5 | 1.2 | | | 1.7 | 3.3 | | 2.0 | 0.4 | 2.5 | №73204, 790-940 ms |

Here measurement error is equal to 10%

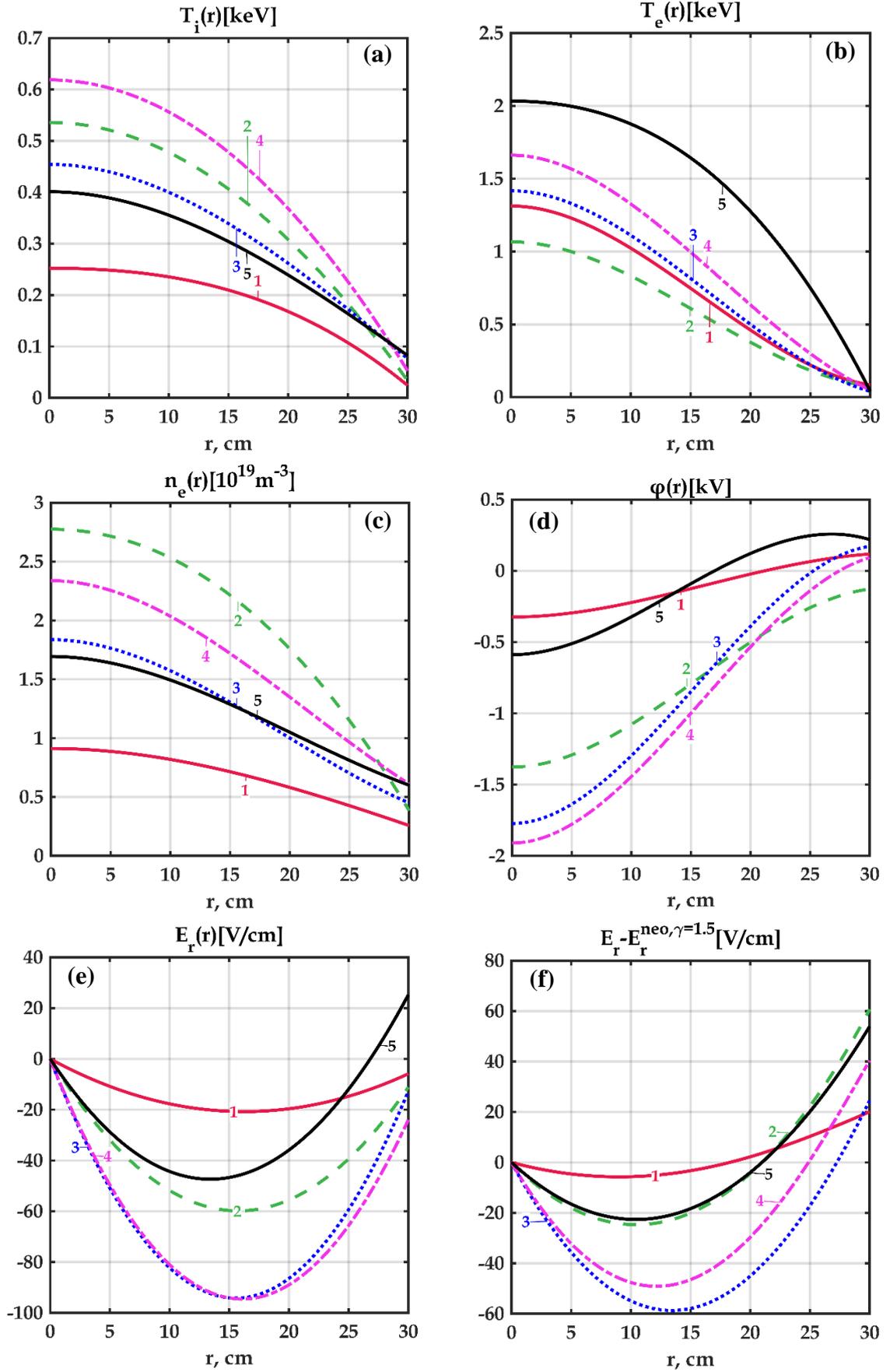

Figure 1. Profiles of plasma parameters: ion temperature (a), electron temperature (b), electron density (c), electric potential (d), radial electric field (c), and the deviation of the measured electric field from the predictions of the neoclassical theory (eq. (2) with $\gamma = 1.5$) (f) in the regimes 1-5 in Table 1.

## 3. Analysis of local correlations

In the analysis of local correlations the strength of the relationship between plasma profiles and the deviation of the measured electric field from neoclassical theory predictions, given by the equation (2), can be estimated using the Pearson correlation coefficient. Pearson's correlation coefficient is a measure of the strength of a statistical relationship between two or more random variables, calculated using the formula:

$$r(x,y) = \frac{\sum_{i=1}^{n}(x_i - \bar{x})(y_i - \bar{y})}{n\sigma_x \sigma_y} = \frac{\sum_{i=1}^{n}(x_i - \bar{x})(y_i - \bar{y})}{\sqrt{\sum_{i=1}^{n}(x_i - \bar{x})^2 \sum_{i=1}^{n}(y_i - \bar{y})^2}}, \quad (3)$$

where x, y are random variables, $\bar{x}$, $\bar{y}$ are mean values of a random variable, n is the number of measurements of a random variable, $\sigma_x$, $\sigma_y$ are standard deviations of random variables. The correlation coefficient varies from -1 to 1. If the correlation coefficient is r=+1, this means that there is a positive correlation between the two values, i.e., if the value of x will increase, then the value of y will also increase (similarly, r=-1 means negative correlation: an increase in x leads to a decrease in y). Correlation coefficient values close to 0 mean weak correlation between values.

Pearson correlation coefficient can be interpreted as a measure of the strength of a linear relationship between two variables. Linear relationship between electron and ion temperature profiles, gradients of these profiles, from the one side, and the residual (1) of the measured electric field and the neoclassical prediction (2), from the opposite side, are shown in figures 2, 3. Pearson's coefficients for correlation between previously mentioned profiles and deviation (1) are shown in figure 4.

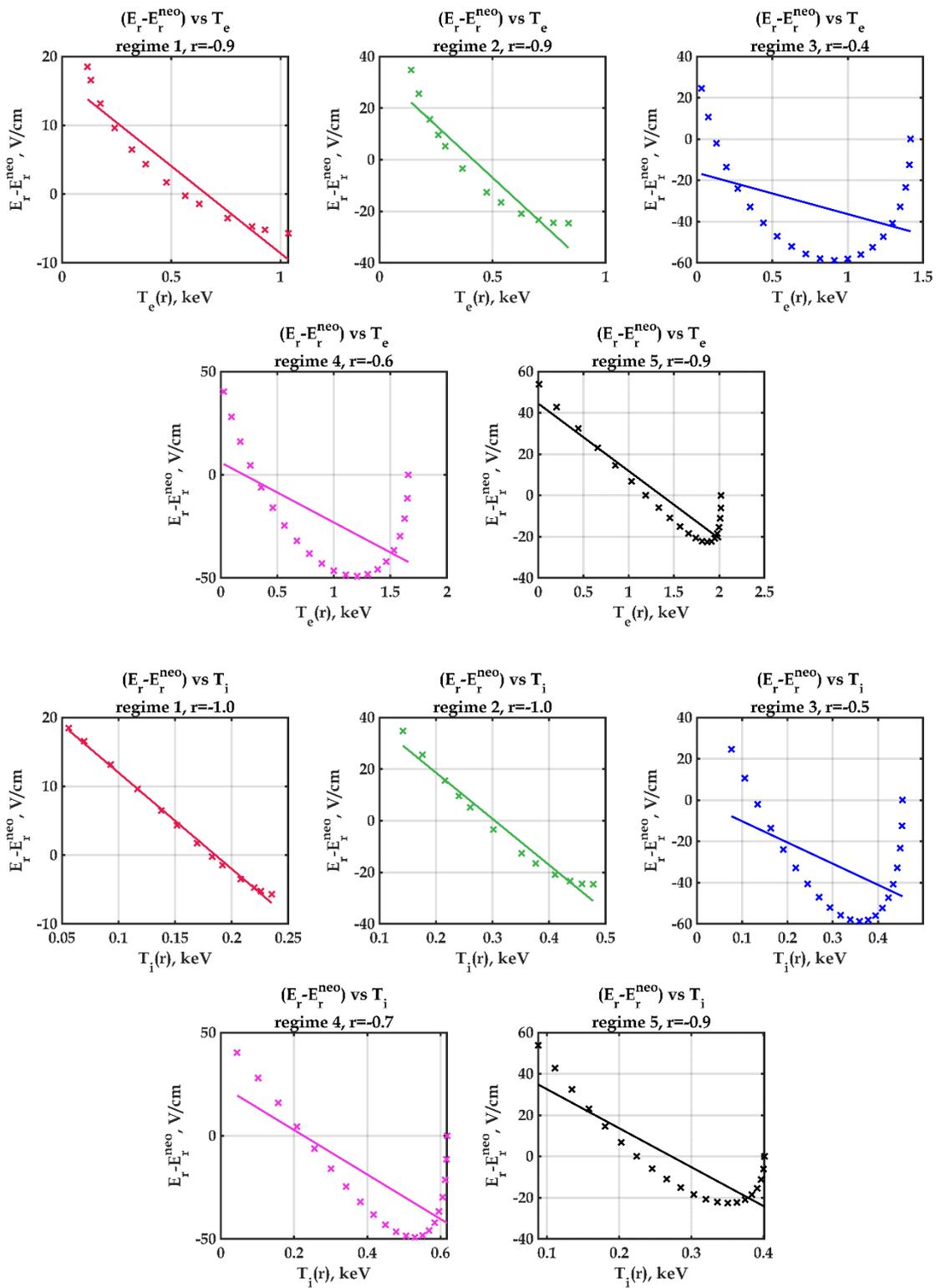

Figure 2. Linear relationship between the temperature profiles (separately for electrons and ions) and the deviation of the measured electric field from the neoclassical prediction (2) for regimes 1-5 in Table 1. Pearson correlation coefficient, indicated in titles, characterizes the strength of a linear relationship.

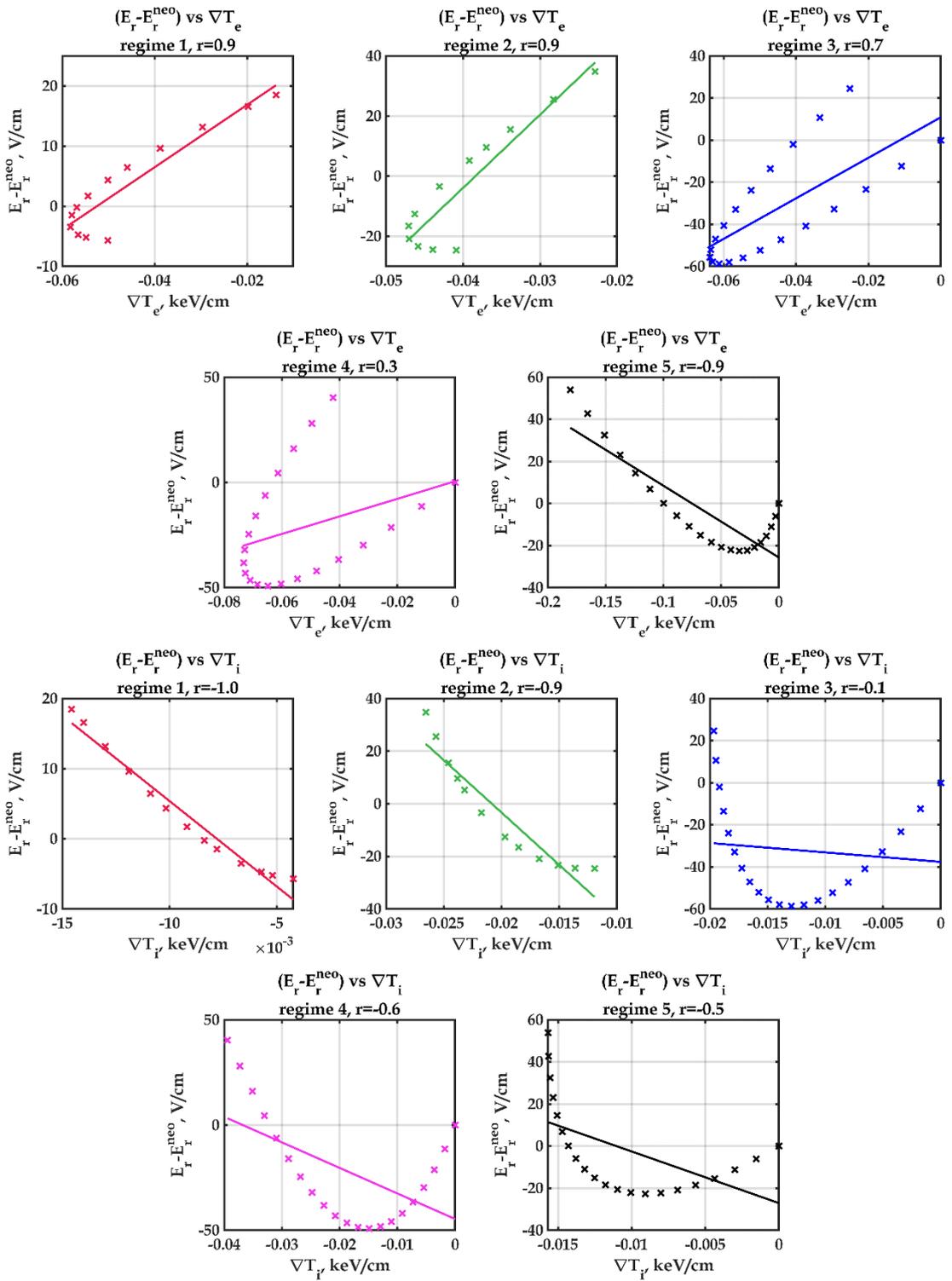

Figure 3. The same as in figure 2, but for the gradients of electron and ion temperature profiles.

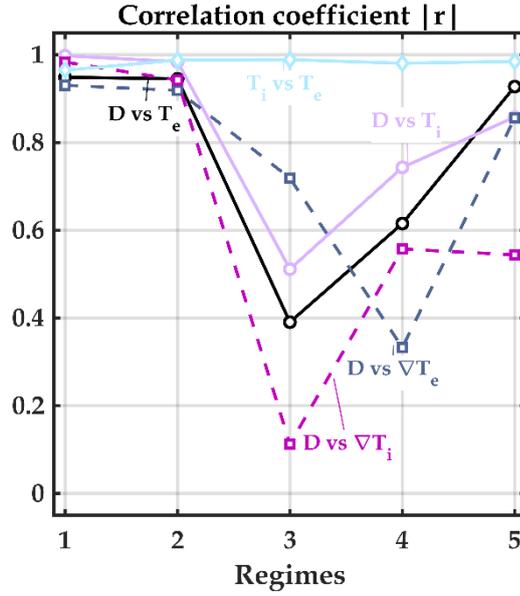

Figure 4. Absolute value of the Pearson's correlation (3) between ion and electron temperature profiles, gradients of these profiles and the deviation $D$ of the measured electric field from the neoclassical prediction (1)-(2), for regimes 1-5 in Table 1.

It can be seen from figures 2-4 that the electrons substantially (not less than ions) contribute to the deviation of the ion equilibrium from the neoclassical theory prediction. Also, it is found that for scenarios 1, 2, 5 there is a strong linear dependence between the ion and electron temperature profiles and the deviation (1), but for scenarios 3-4 the linear dependence does not work. Linear relationship between the gradients of ion and electron temperature profiles and the deviation (1) is generally weaker than linear relationship between deviation (1) and profiles $T_e$ and $T_i$. These results suggest that it's worth to make a step in the opposite direction, in the sense of the functional analysis, and look for a nonlocal, integral type of the correlation between the deviation (1) and the profiles of ion and electron temperature.

## 4. Analysis of nonlocal correlations

Nonlocal correlations can be treated with an inverse problem under the assumption of an integral equation relationship between the deviation (1) and $T_e$ and $T_i$ profiles The inverse problem assumes a search for the functions $G_e$ and $G_i$, which enter the following relationship between the known functions $D(r,t)$, $T_e(\rho,t)$, and $T_i(\rho,t)$:

$$D(r,t) = \int_0^1 G_e(r,\rho,t,\mathbf{T_e(\rho,t)}) \cdot T_e(\rho,t) d\rho + \int_0^1 G_i(r,\rho,t,\mathbf{T_i(\rho,t)}) \cdot T_i(\rho,t) d\rho, \quad (4)$$

where $r$ is the minor radius of toroidal plasma, $\rho$ is the normalized (dimensionless) minor radius.

Here, we restrict ourselves to analyzing the correlation with temperature profiles while a dependence on the density profile is not considered.

In equation (4), the volume integral is assumed, however the respective Jacobian can be included into the unknown functions $G_e$ and $G_i$. In the simplest linear response approximation, we get:

$$D(r,t) = \int_0^1 G_e(r,\rho) \cdot T_e(\rho,t) d\rho + \int_0^1 G_i(r,\rho) \cdot T_i(\rho,t) d\rho. \quad (5)$$

In this case, the inverse problem of multi-objective optimization has the form:

$$\min_{G_e(r,\rho),G_i(r,\rho)} \left( D(r_i,t_1)^2, D(r_i,t_2)^2, ..., D(r_i,t_N)^2 \right) \quad (6)$$

This problem can be reduced to a single-objective optimization problem:

$$\min_{G_e(r,\rho),G_i(r,\rho)} \left( \sum_{i,j} D(r_i,t_j)^2 \right) \quad (7)$$

The presentation of the residual (1) in the form of integral operators assumes the dominance of non-local mechanisms in generating the deviation of the ion equilibrium from the neoclassical theory. In the case of reducing the functions $G_e$ and $G_i$ to local operators (close to the Dirac delta functions), this will mean the locality of the equilibrium equation. The separation of the ionic and electronic contributions in (5) is impute and is aimed at comparing the separate contributions of ions and electrons to the deviation (1).

The inverse problem (5), (7) is solved on the $r$ and $\rho$ grids with 5 and 10 nodes, respectively, where $r$ stands for the effective minor radius of the toroidal plasma as an argument of the function in the left hand side of (4), while $\rho$ stands for the normalized (dimensionless) minor radius as a dumb variable in the integrals in the right hand side of (4). Each of the discrete functions $G_i(r, \rho)$ and $G_e(r, \rho)$ contains 50 unknown values that need to be found within the framework of solving the inverse problem (5), (7). For each value of $r$ on a new grid we have only 5 equations for the residual (1), obtained from experimental regimes 1-5 (Table 1). To make the inversed problem less ill-posed, we generate additional 100 experimental regimes, obtained from the experimental regimes 1-5 by adding a random error of 5% to the profiles $T_e$, $T_i$ and plasma electric potential $\varphi$. The resulting inverse problem (5), (7) is solved with the Matlab Genetic Algorithm with enabled hybrid function (fmincon), that continues the optimization after the original Genetic Algorithm solver terminates. Figures 5 and 6 show the solution of the inverse problem (5), (7).

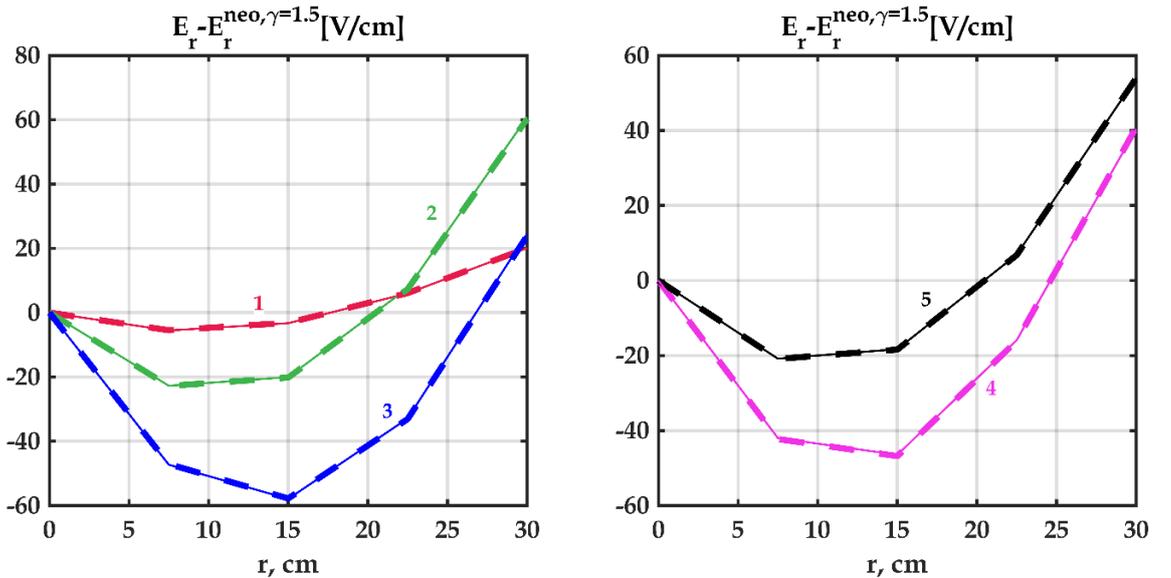

Figure: 5. Results of solving the inverse problem (5), (7) for regimes 1-5 in Table 1: (a) ohmic regimes 1-3, (b) regimes with ECRH. Solid thin lines present experimental data, dashed lines show solution of the inverse problem (5), (7).

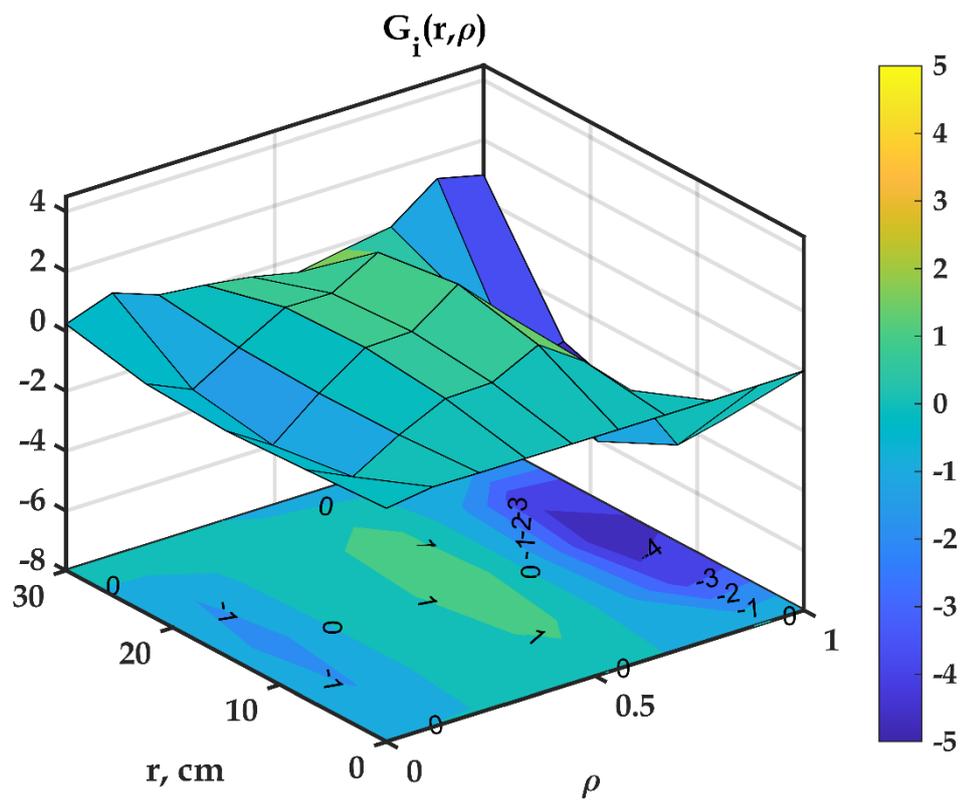

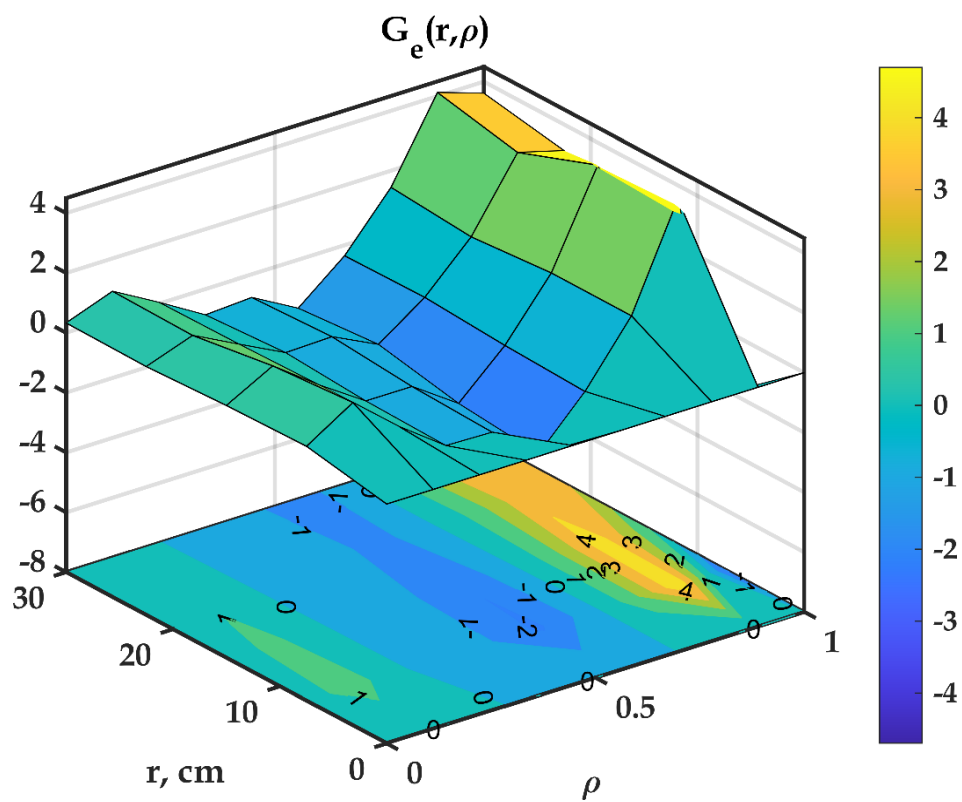

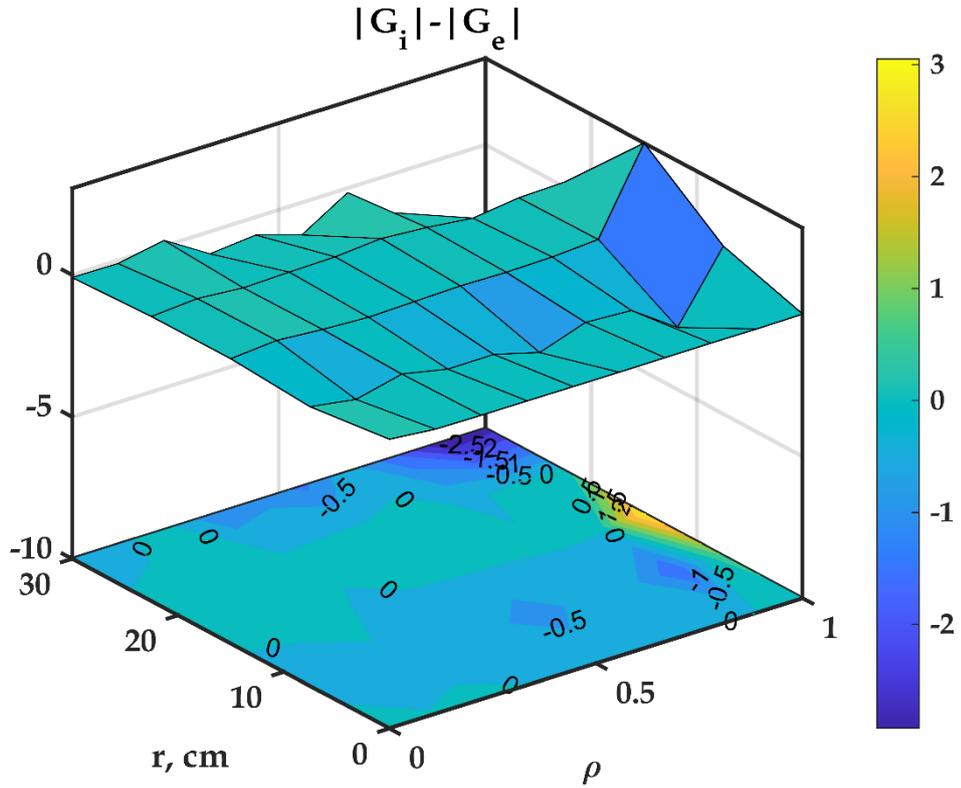

Figure 6. Functions $G_e$ and $G_i$ (in arb. units), defined in (5) and obtained by solving the inverse problem, for regimes 1-5 in Table 1.

It is seen from figure 6 that in the essential part of the plasma the contribution on ions and electrons to the deviation (1) is of the same order of magnitude and has the opposite sign while in the peripheral plasma the contribution of the electrons dominates over that of ions. On the whole, the contribution of ions and electrons to the deviation (1) for the force equilibrium of ions are comparable, and the electrons contribute not less than the ions.

## 5. Discussion and conclusions

We present the results of the analysis of the deviation of the force equilibrium for ions from the neoclassical theory prediction, calculated using the direct measurements of the radial electric field, in the view of its possible local and nonlocal correlation with the profiles of electron, $T_e$, and ion, $T_i$, temperature in the T-10 tokamak. The data for discharges with zero, weak and strong auxiliary heating (electron cyclotron resonance heating) are analyzed. One can draw the following conclusions.

1. Analysis of local correlations is carried out by means of the Pearson's correlation. It is shown that for some regimes of the T-10 tokamak operation there is a strong linear dependence between the ion and electron temperature profiles and the deviation (1), however for another regimes linear dependence does not work. Linear relationship between the gradients of ion and electron temperature profiles and the deviation (1) is generally weaker. These results suggest that it's worth to look for a nonlocal, integral type of the correlation between the deviation (1) and the profiles of ion and electron temperature.

2. Analysis of possible nonlocal correlations is treated with an inverse problem under the assumption of an integral equation relationship between the deviation and $T_e$ and $T_i$ profiles. It is found that in the essential part of the plasma the contribution on ions and electrons to the deviation (1) is of the same order of magnitude and has the opposite sign while in the peripheral plasma the contribution of the electrons dominates over that of ions.

3. Overall, the electrons substantially (not less than ions) contribute to the deviation of the force equilibrium for ions from the neoclassical theory prediction both in the local and nonlocal correlation models.

The latter conclusion is in a qualitative agreement with the hypothesis [12] for the role of two-fluid effect in plasmas with strong electric current (Hall effect), which leads to a strong contribution of electron dynamics to the formation of the radial electric field in the central part of the plasma column.

**Acknowledgements**

This work was supported in part by the Russian Foundation for Basic Research (project 18-07-01269-a). The T-10 tokamak experiments were supported by the Russian Science Foundation, Project 19-12-00312. A.V.M. and A.B.K. were partly supported by Competitiveness program of NRNU MEPhI.